\documentclass[11pt,a4paper]{article}
\usepackage{jinstpub}
\usepackage{amsmath}
\usepackage{graphicx}
\RequirePackage[pagewise,mathlines]{lineno}

\begin{document}
\title{Using precision timing to improve particle tracking}

\author{Authors: Spencer R. Klein}
\emailAdd{srklein@lbl.gov}
\affiliation{Nuclear Science Division, Lawrence Berkeley National Laboratory, 1 Cyclotron Road, Berkeley, CA 94720 USA}
\date{\today}
\abstract{Silicon tracking detectors provide excellent spatial resolution, and so can provide excellent momentum resolution for energetic charged particles, even in compact detectors.  However, at lower momenta, multiple scattering in the silicon degrades the momentum resolution.  We present an alternate method to measure  momentum and alleviate this degradation, using  silicon detectors that also incorporate timing measurements.  By using timing information between two silicon layers, it is possible to solve for the the radius of curvature, and hence the particle momentum, independent of multiple scattering within the silicon.  We consider three examples: an all-silicon central tracker for an electron-ion collider, a simplified version of the CMS detector, and a forward detector for an electron-ion collider.  For a 75 cm diameter tracker in a 1.5 T magnetic field, timing can improve the momentum determination for particles with momentum below 500 MeV/c.  In the 3.8 T CMS magnetic field and 1.2 m radius tracker, timing can improve tracking up to momenta of 1.3 GeV/c.  The resolution is best at mid-rapidity.   We also discuss a simpler system, consisting of a single timing detector outside an all-silicon tracker.}
\maketitle

\section{Introduction}

Most high-energy particle and nuclear physics detectors use position-sensitive detectors to track charged particles as they move through a (usually) solenoidal magnetic field.  From these space points, the track is reconstructed as a helix, with transverse momentum ($p_T$) proportional to the bending radius.  At high $p_T$, the particles bend only slightly, so their sagitta is small, and the momentum resolution is determined by the precision of position measurements.   At lower momentum, though, the momentum determination is limited by multiple scattering in the detectors.   Good momentum resolution leads to good mass resolution, which is important in reducing the backgrounds in decays like $\Lambda\rightarrow p\pi^-$, and $D^{*+}\rightarrow D^0\pi^+$.  Both of these final states include low-momentum daughter particles, and will be important at a future electron-ion collider \cite{Accardi:2012qut}. 

Silicon active pixel detectors have excellent position resolution (typically in the 10 $\mu$m to 30 $\mu$m range), and so offer excellent momentum resolution.  They are used as primary trackers at the Large Hadron Collider and are being considered as primary detectors for a future electron-ion collider.   However, these detectors are relatively thick: $50-100\ \mu m$ in current devices, with the potential for a factor of two reduction by using thinned silicon detectors.   These detectors must also be mounted on support structures and provided with power and cooling, and a means to read out the data.  Current state of the art detectors are as thin as 0.3\% radiation length ($X_0$) \cite{Contin:2017mck,Greiner:2019ewb}.  A planned replacement for the ALICE ITS inner layers could be considerably thinner, 0.05\% $X_0$,  for small-area inner layers \cite{ITS3}.  The larger outer layers will need to be thicker to be mechanically stable.  Since multiple scattering scales as the square root of the thickness, these thinner detectors ofter significant improvements, but are not decisive in reducing multiple scattering. 

This paper will present a technique to compensate for multiple scattering, and thereby improve the momentum resolution of silicon tracking detectors, by using precision timing to determine the flight time between silicon detector layers.  This timing is available in new detector designs, which are capable of achieving excellent timing resolution, below 30 psec \cite{Sola:2017zty,Cartiglia:2019fkq}.  Larger multilayer silicon pad detectors have achieved resolutions below 20 psec \cite{Cartiglia:2016voy}.  The technology of fast timing in silicon detectors is nicely reviewed in Ref. \cite{Sadrozinski:2017qpv}. A timing resolution of 12 psec could be achievable for a heavily biased (200 V) 50\ $\mu$m thick sensor with a very low noise readout  \cite{Riegler:2017xbh}; sub 10-psec resolution could be achievable with slightly thinner sensors. The 10 psec level matches a planned silicon time-of-flight system for an EIC detector \cite{Repond:2019hth}, so we will adopt it here.   

This paper will consider two cases: a possible all-silicon detector for an electron-ion collider, and for the CMS detector at the LHC.  It should be noted that the the CMS, ATLAS and LHCb detectors are all already considering adding timing capability for high-luminosity LHC running, in order to better associate hits and tracks with individual vertices.  

\section{An EIC tracking detector}
\label{sec:det}

We consider an all-silicon central detector in a uniform solenoidal magnetic field.   If the magnetic field is not uniform, the calculations are more involved.   This model detector is tailored to fit in the 1.5 T magnet used by sPHENIX, leaving room for particle identification devices plus some calorimetry, {\it i. e.}, so it is the same size as the sPHENIX tracker  \cite{Roland:2019cwl}, with an assumed 75 cm outer radius.  The silicon is based on the ALICE inner tracking system (ITS) detector \cite{Abelevetal:2014dna}, but with the seven layers  distributed over a wider radial range, from 2.2 cm to 75 cm in radius.  The lengths were adjusted to roughly cover the pseudorapidity range $|\eta| \le1$, as listed in Tab. \ref{tab:silicon}. The layer thicknesses were adjusted from the ITS values by assuming that the detectors would be air cooled, as with the STAR Heavy Flavor Tracker (HFT) \cite{Contin:2017mck}, and with the silicon thinned to 50\ $\mu$m.  The division into three thinner inner layers and four thicker outer layers with larger pixels and thicker supports was retained.  The thicker supports are needed to handle the longer staves. 

\begin{table}

\begin{tabular}{lrrrr}
Layer & Radius & Length &Thickness & Pixel size \\
 & (cm) & (cm) & ($X_0$) & ($\mu m\times\mu m)$ \\
\hline
1 & 2.2 & 27 &0.2\% & $25\times50$ \\
2 & 3.1 & 27 &0.2\% & $25\times50$ \\
3 & 4.0 & 27 & 0.2\% & $25\times50$ \\
4 & 12.0 & 72 & 0.6\% & $50\times100$ \\
5 & 20.0 & 200 & 0.6\% & $50\times100$ \\
6 & 50.0 & 200 & 0.6\% & $50\times100$ \\
7 & 75.0 & 200 & 0.6\% & $50\times100$  \\
\hline
\end{tabular}
\caption{Key dimensions for an all-silicon tracking detector for an electron-ion collider.   The silicon characteristics are derived from the ALICE ITS design, but with thinned silicon and air cooling, instead of water.}
\label{tab:silicon}
\end{table}

\begin{figure}
\center{\includegraphics[width=3in]{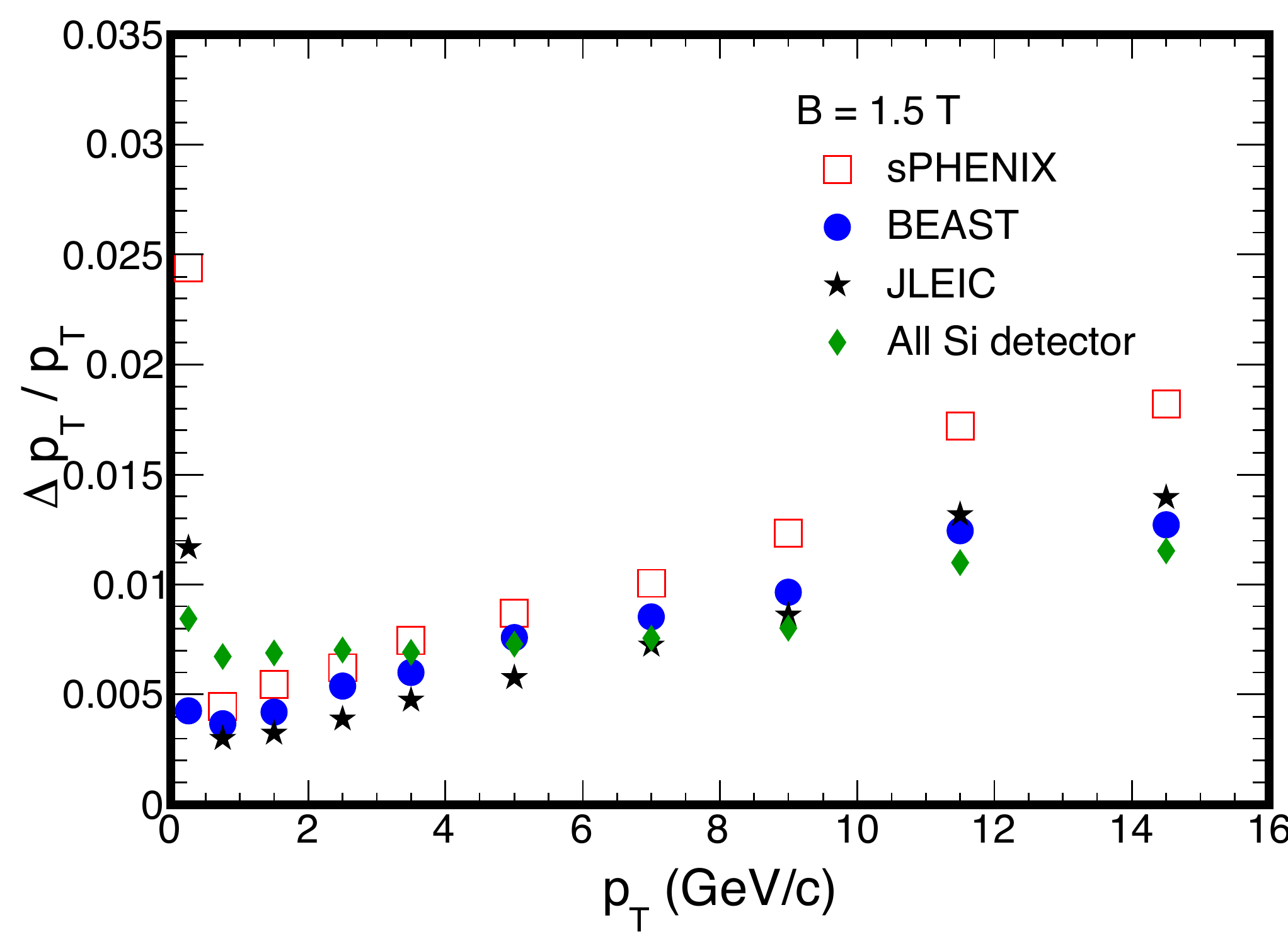}}
\caption{Momentum resolution $\Delta p/p$ for the all-silicon detector design described in Tab. \ref{tab:silicon}, compared with parallel calculations of the momentum resolution of several other detectors proposed for an EIC. }
\label{fig:resolution}
\end{figure}

The no-timing tracking resolution was determined with a simple model,  that used the silicon position resolution and multiple scattering for isolated tracks {\it i. e.} assuming perfect tracking.    Figure \ref{fig:resolution} shows the momentum resolution at pseudorapidity ($\eta=0$), compared with three other proposed EIC detectors.  The other detectors primarily use  gaseous detectors for momentum resolution: sPHENIX \cite{Roland:2019cwl}, BEAST \cite{Fazio:2017vxq}, and JLEIC \cite{Morozov:2019uza}.  The resolutions were calculated in the same framework, ensuring that this is an apples-to-apples comparison between silicon and gaseous detectors.  

The silicon detector has excellent performance at high momentum, because of the excellent spatial resolution.  However, at low momenta, it's resolution degrades rapidly because of the higher detector thickness.   The resolution may be parameterized as a quadrature sum
\begin{equation}
\frac{\Delta p}{p} \approx 0.01 p \otimes \frac{0.025}{p}
\label{eq:res}
\end{equation}
where p is in GeV/c.  The first term accounts for the precision of measuring the sagitta, and the second is due to multiple scattering.   The first term is smaller for an all-silicon detector than for a comparably sized gas detector because of the superb space-point resolution.   The second term is larger for silicon than for a comparable gas detector because the silicon detector contains more material.  For momenta below about 1 GeV/c, the momentum uncertainty is predominantly from the multiple scattering.  The second term in eq. \ref{eq:res} scales roughly as the square root of the thickness.   A reduction in layer thickness moves the cross-over point, but such a detector will still struggle at low momentum.

\section{Resolution with timing}

We now consider the how timing could improve the tracking performance of this silicon detector.    We assume that the particle identity is known, so we can relate momentum to velocity.  The soft particles considered here can be identified using time-of-flight based on a track curvature reconstructed without the use of timing.  

\begin{figure}
\center{\includegraphics[width=3in]{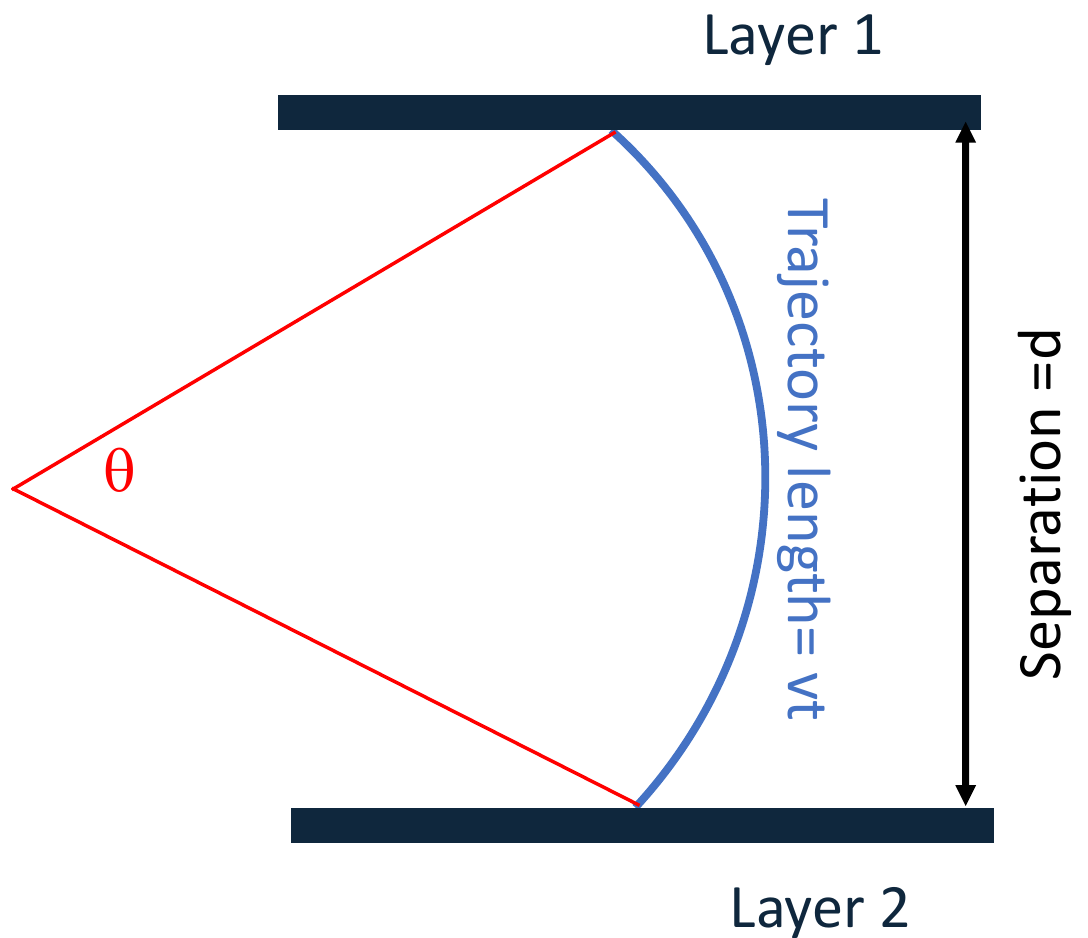}}
\caption{The geometry for a two-layer detector.   The particle covers a path length $L=R\theta$, where the radius of curvature $R$ is  inversely related to the momentum.}
\label{fig:geometry}
\end{figure}

We initially consider a single pair of silicon detectors, with separation $d$, at $y=0$ as is shown in figure \ref{fig:geometry}.  Between the layers, charged particles follow an arc with radius of curvature $R =  \alpha B p_T$  where $B=1.5$ T is the solenoidal magnetic field and $\alpha =$ 300 MeV/T.      The arc length is $L=R\theta$, and the particle moving at velocity $v$ takes time $t=L/v$ to travel between the two layers.    The separation vector $d$ is a chord for the circle shown, with 
\begin{equation}
d=2R \sin{\big(\frac{\theta}{2}\big)}
\end{equation}
By combining these two equations, we can eliminate $\theta$:
\begin{equation}
\frac{d}{2R} = \sin{\big(\frac{vt}{2R}\big)}
\label{eq:big}
\end{equation}
Equation \ref{eq:big} is analytically intractable, but  easily solved numerically.
In the limit of large momentum, $R\rightarrow\infty$ and  $d=vt$.    The velocity $v= p/\sqrt{p^2+m^2}$, where $m$ is the particle mass.  This approach works as long as the particles do not curl up, so $\theta<\pi$ ($R>d/2$), {\it i. e.} $p_T>d/(2\alpha B)$.  

Multi-layer detectors can be treated as a combination of two-plane systems.  In a simple case, three planes with two separations, the time at the intermediate layer completely cancels out as long  there is no energy loss (only multiple scattering) in the intermediate layer.  The momentum can be determined using the timing in the two outer layers, and the total path length.  The path length is the sum of the two arc lengths; any change in direction due to scattering in the intermediate layer is immaterial as long as the intermediate point is used to find the total path length for the two arcs.  

For some geometries, the cancellation of the intermediate times may be incomplete.   This is seen in another example: two pairs of closely spaced layers, with a big separation between the pair.  The path length is dominated by intermediate arc, and all four detectors can contribute to measuring that time.   Here, we conservatively assume that the timing from the inner layers completely cancels, and use the total time, determined from the inner and outer detectors to determine the momentum.  

We consider two scenarios.  The first is an isolated track, where we use the inner and outer silicon layers to determine the flight time.  The second is an event consisting of many tracks.  For these events, we assume that the interaction time can be well constrained, with uncertainty less than the single-layer silicon timing,  using beam-beam counters and silicon timing from many tracks.  These tracks should be particle-identified, to avoid ambiguity regarding their velocity; an iterative procedure may be required to progressively fit the interaction time and particle identities.  With this, the resolution depends only on the outer timing layer and the resolution improves by $\sqrt{2}$, as long as the innermost position measurement is close enough to the beampipe (so that the the beampipe and inner layer can be treated as a single scatterer).  Since thinned silicon can be bent to fit around the beampipe, this is technically feasible \cite{Burghartz}.

Away from $y=0$,  eq. \ref{eq:big} must be modified.  The pseudorapidity introduces a displacement $\Delta z$ along the beam direction (in $z$). $\Delta z$ is determined from the radii of the inner ($R_{\rm min}$) and outer ($R_{\rm max}$) silicon layers.
\begin{equation}
 \Delta z=(R_{\rm max}-R_{\rm min})\sinh(\eta).
 \end{equation}
 The pathlength (starting at the origin) is 
 \begin{equation}
vt=\sqrt{(R\theta)^2+\Delta z^2}.
 \end{equation}
so
 \begin{equation}
 \theta=\sqrt{\frac{v^2t^2-\Delta z^2}{R^2}}
\end{equation}
The equation for the chord (in the radial direction) is unchanged, so
\begin{equation}
\frac{d}{2R} = \sin(\frac{\theta}{2}) = \sin\big(\frac{\sqrt{v^2t^2-\Delta z^2}}{2R}\big)
\end{equation}
which is more complicated, but, since we're already solving this numerically, doesn't add additional complications.  

\section{Results}

We first consider the detector described in Tab. \ref{tab:silicon}, assuming a silicon time resolution of 10 psec, for two scenarios.
The first is an isolated track, where no other timing information is available. The second scenario assumes that the time of the primary vertex can be accurately determined, with precision $\sigma \ll 10$ psec.   In the former, innermost and outermost silicon layers each contribute their timing uncertainty, while in the former, the primary vertex provides a highly accurate initial time, so only the resolution of the outer layer contributes.   The primary vertex timing improves the $vt$ resolution by $\sqrt{2}$.  It is also necessary to consider multiple scattering in the beampipe, which is thicker than the silicon layers. We assume that the first detector layer is mounted on the beampipe; this is possible with thinned silicon, which can be bent to a suitable radius \cite{Burghartz}. 

We consider pions and protons with momenta from about 165 MeV/c to 600 MeV/c.  The lower momentum limit is set by the requirement that the tracks reach the outer layer, rather than curling up.  The resolution degrades at higher momentum, and conventional tracking approaches performs much better.   Electrons are likely to lose energy via bremsstrahlung in the silicon, so we do not consider them. For each momentum, we calculate the travel time between each layer.  Then, we alter this timing by $\pm 1 \sigma$ (10 psec) and use eq. \ref{eq:big} to determine the measured momentum.  

\begin{figure}
\center{\includegraphics[width=3.25in]{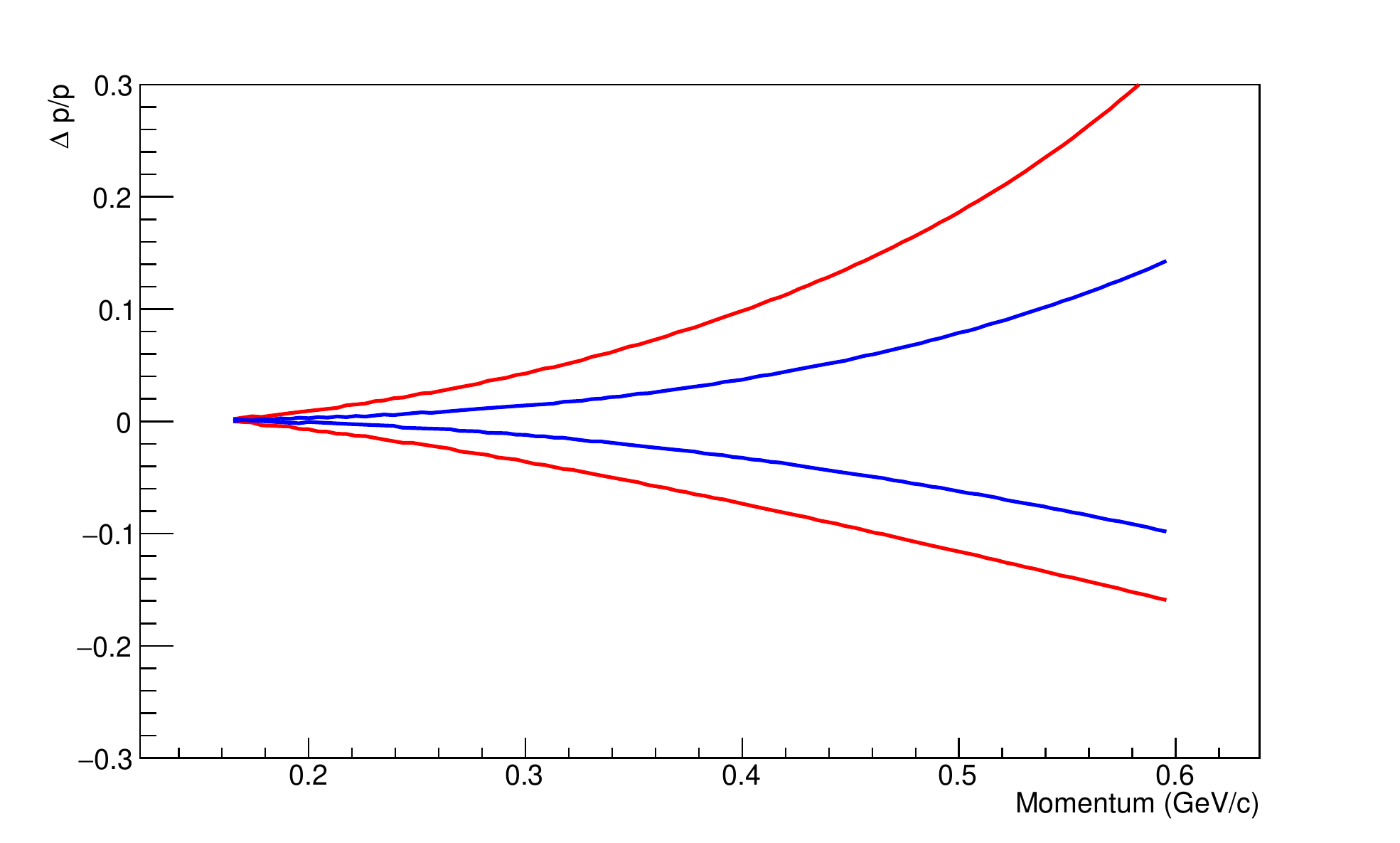}}
\caption{The momentum resolution at $\eta=0$ for the EIC detector in Table \ref{tab:silicon} for a timing resolution of 10 psec for isolated tracks.  The red curves show the $\pm 1\sigma$ resolution for pions, while the blue curves are $\pm 1\sigma$ for protons.}
\label{fig:twodetector}
\end{figure}

Figure \ref{fig:twodetector} shows the resolution for isolated pions and protons.   The problem is non-linear, so the resolution is asymmetric. 
The resolution is significantly better for protons, because their velocity is lower, so their flight time is longer and the relative timing resolution $\Delta T/T$ improves.   Slower particle may also deposit more energy in the silicon layers, potentially improving the timing resolution.  However, we do not consider that factor here.  

The calculated resolution is zero at the minimum momentum, when $R\rightarrow d/2$, and $\theta\rightarrow\pi$.  The particle is moving perpendicular to the chord $d$ at both ends of its path, so $dR/dL\rightarrow 0$ and the resolution becomes ideal.  A more detailed treatment would include contributions due to positional uncertainty and scattering in the inter-silicon gas.

165 MeV/c  is the minimum momentum for the particles to reach the outer layer.  The use of intermediate layers would allow us to extend the treatment down to lower momentum.  The overall minimum momentum is set by the number of silicon layers required to reconstruct the track, likely three layers.  Per Tab. \ref{tab:silicon}, this implies momentum resolution down to $\approx 10$ MeV/c.  This is, of course, unrealistic, since it does not include $dE/dx$ energy loss in the detector; furthermore, the large multiple scattering at these low energies would greatly complicate pattern recognition. 
 
\begin{figure}
\center{\includegraphics[width=3.25in]{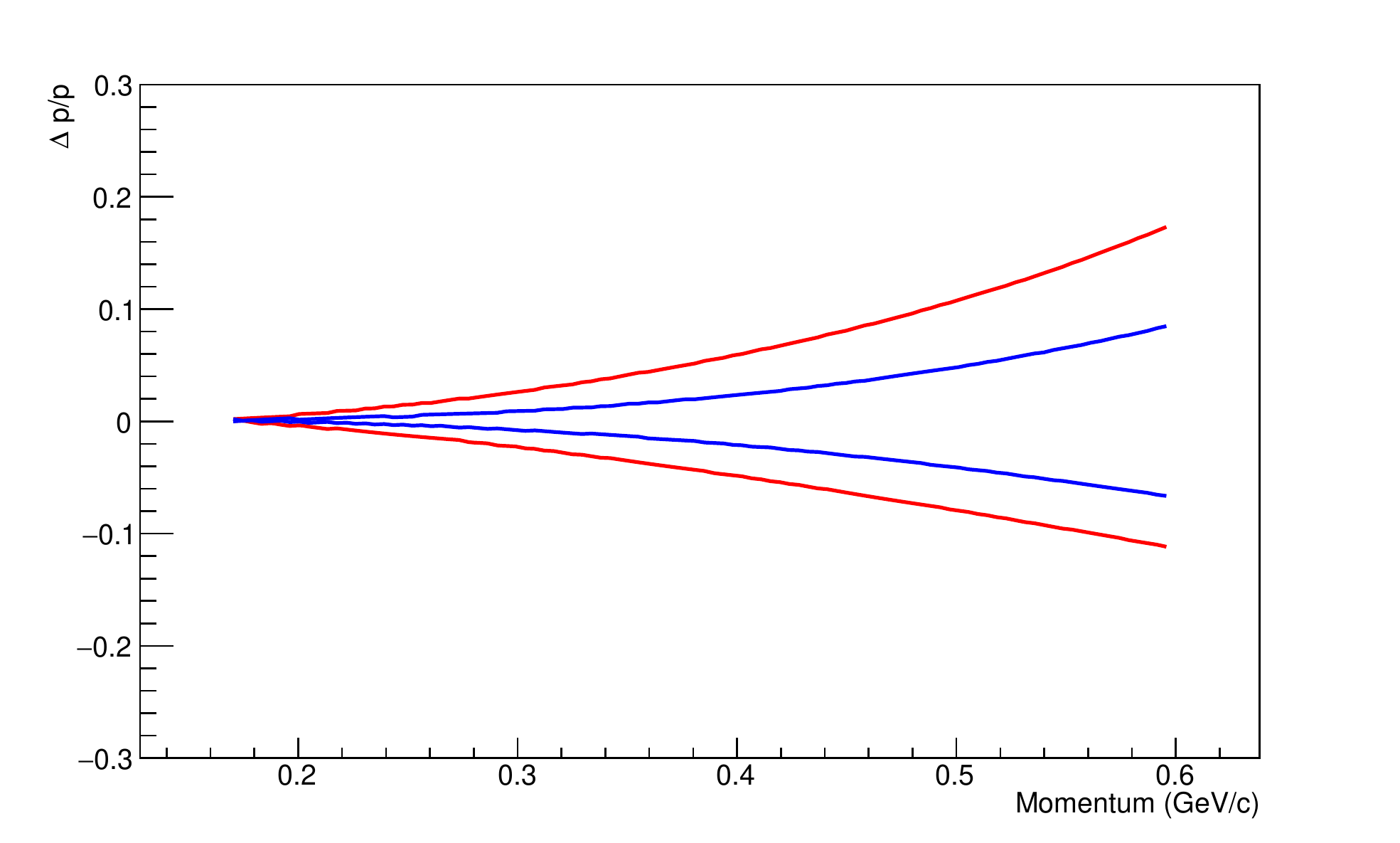}}
\caption{The momentum resolution at $\eta=0$ for the EIC detector in Table \ref{tab:silicon} for a timing resolution of 10 psec for higher multiplicity events, where the vertex position and interaction time is well known.  The red curves show the $\pm 1\sigma$ resolution for pions, while the blue curves are $\pm 1\sigma$ for protons.}
\label{fig:vertexplatedetector}
\end{figure}

 Figure \ref{fig:vertexplatedetector} shows the resolution for higher multiplicity events, where the primary vertex time is well measured.  The timing resolution is $\sqrt{2}$ better, and the resolution improves commensurately.  
 
\begin{figure}
\center{\includegraphics[width=3.25in]{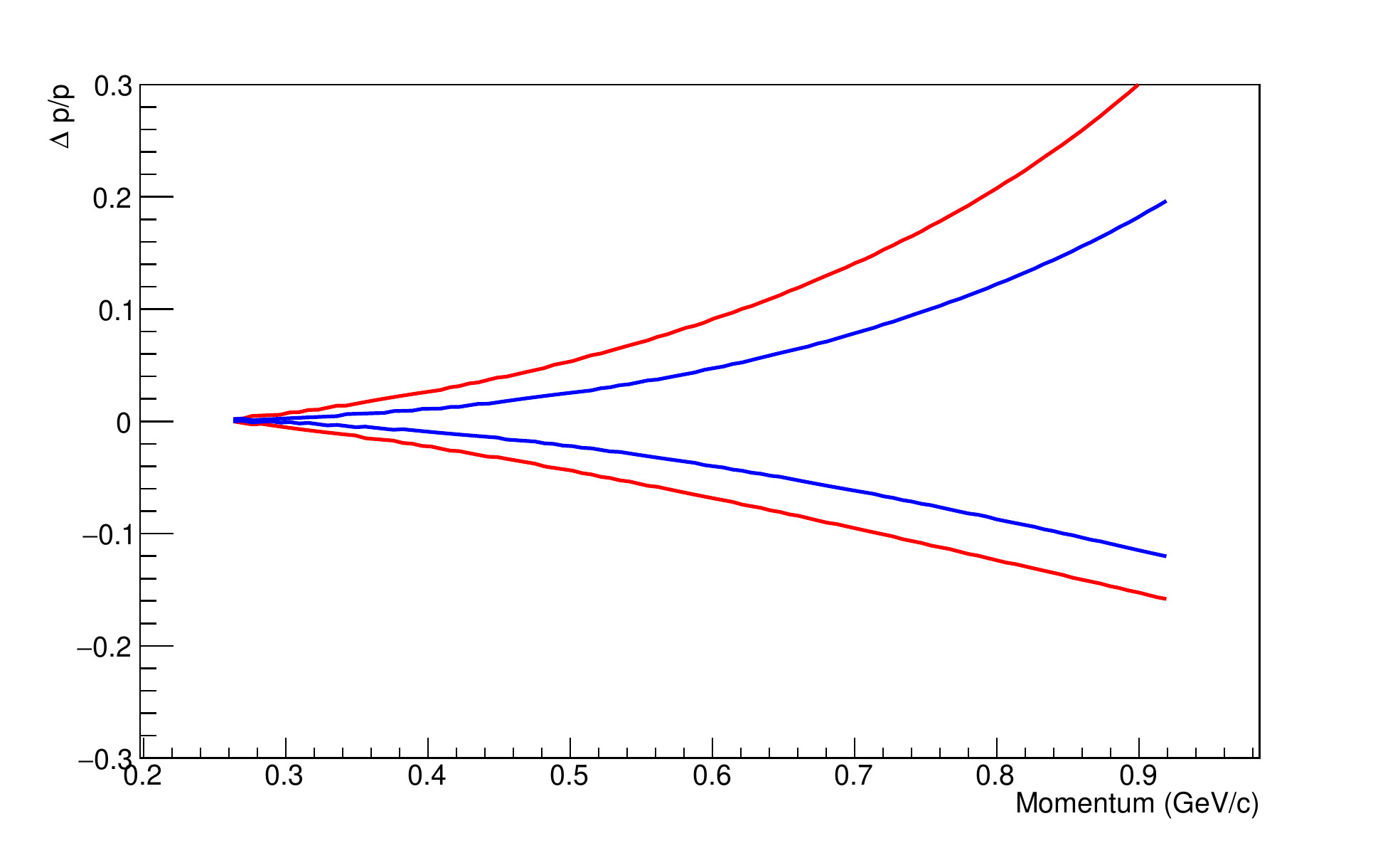}}
\caption{The total momentum resolution at $|\eta|=1$ for the EIC detector in Table \ref{tab:silicon} for a timing resolution of 10 psec for events for which the vertex time is accurately found.  The red curves show the $\pm 1\sigma$ resolution for pions, while the blue curves are $\pm1\sigma$  for protons.}
\label{fig:forward}
\end{figure}

Figure \ref{fig:forward} shows the total $p$ (not $p_T$) momentum resolution for the same detector  at $|\eta|=1$, again assuming that the vertex position is known.  The minimum momentum is higher by a factor $\cosh(|\eta|)$.    Significantly above threshold, the momentum resolution is similar.  The $p_T$ is a factor of $1/\cosh(|\eta|)$ smaller, so the resolution is worse than in figure \ref{fig:vertexplatedetector}.    For $|\eta|=1$, the degradation from the forward trajectory nearly cancels out the improvement from the known vertex timing, and the $p_T$ resolution is very similar to figure \ref{fig:twodetector}.

\begin{figure}
\center{\includegraphics[width=3.25in]{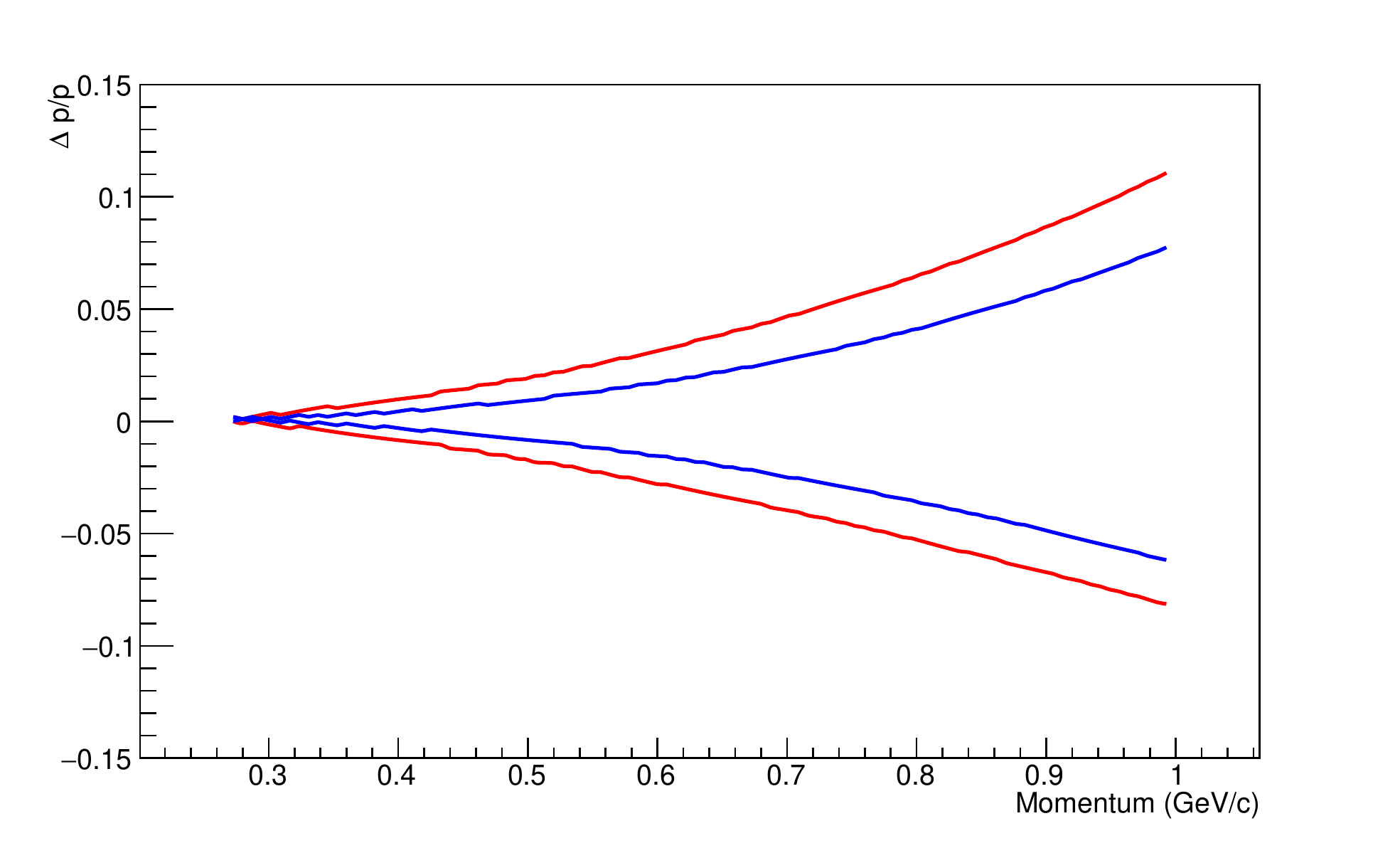}}
\caption{The momentum resolution at $\eta=0$ for a detector with 120 cm outer radius, with a timing resolution of 10 psec for events for which the vertex time is accurately found.  The red curves show the $\pm1\sigma$ resolution for pions, while the blue curves are $\pm1\sigma$ for protons.  The y axis is expanded from the previous plots.}
\label{fig:detector12m}
\end{figure}

The resolution will improve with increasing path length and with increasing magnetic field.  Figure \ref{fig:detector12m} shows the resolution for a detector in the same 1.5 T magnetic field, but with a 120 cm outer radius.  The resolution is significantly improved.  

\begin{figure}
\center{\includegraphics[width=3.25in]{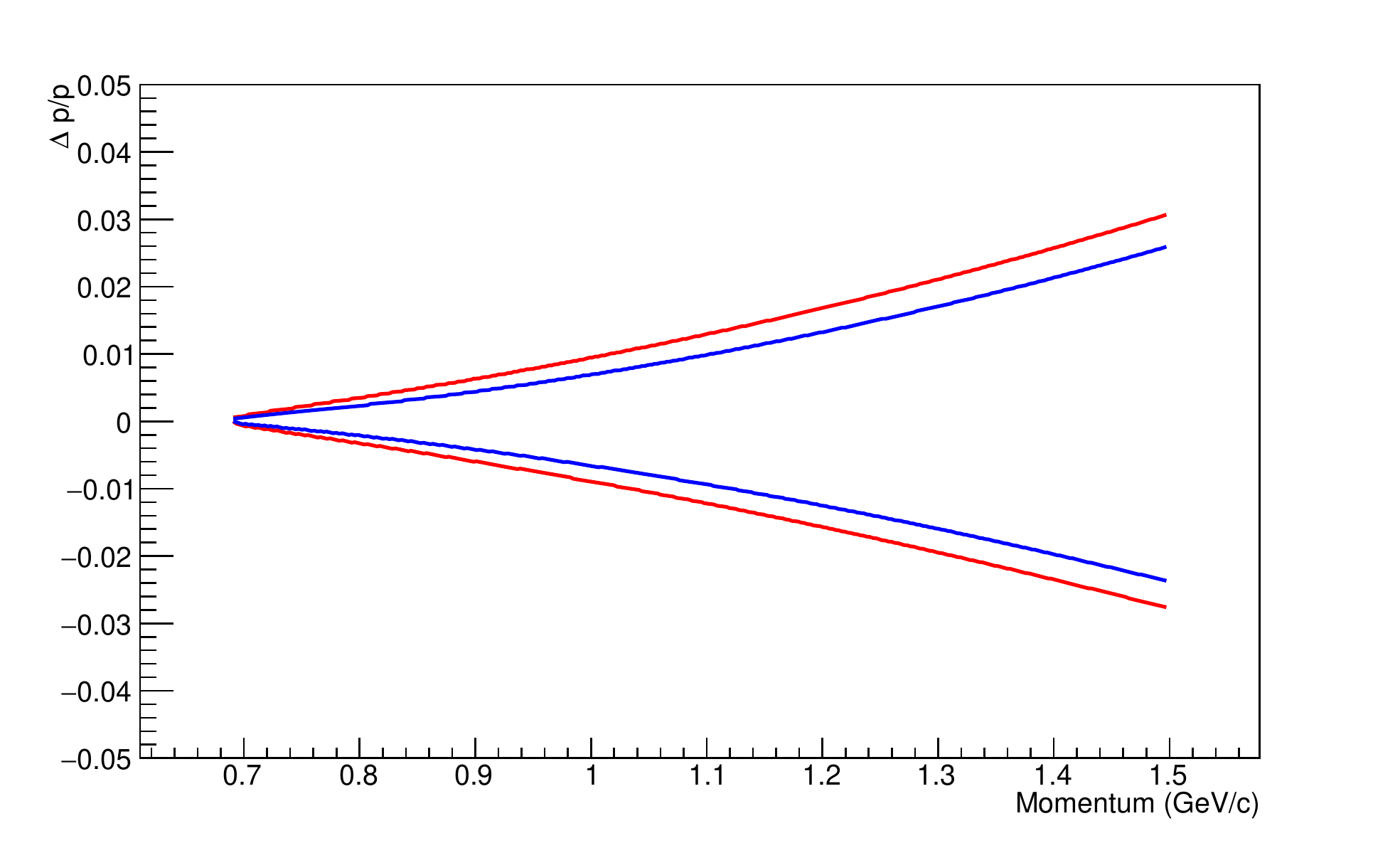}}
\caption{The momentum resolution at $\eta=0$ for a CMS-like detector for a timing resolution of 10 psec for events for which the vertex time is accurately found.  The red curves show the $\pm1\sigma$ resolution for pions, while the blue curves are $\pm1\sigma$ for protons.}
\label{fig:CMS}
\end{figure}

We now turn to a detector modeled after CMS, with the same 1.2 m radius solenoidal tracking system but in a 3.8 T field.   The resolution for this case is shown in figure \ref{fig:CMS}, assuming that the primary vertex timing is well known.   The resolution is improved compared to figure \ref{fig:detector12m}, with the improvement proportional to the ratio of the magnetic fields, as expected. 

\section{A partial timing detector}

In the model for events with well reconstructed vertices, only the outermost timing measurement is used; timing on all layers is not needed.  Although the timing resolution is very important, in most applications, the multiplicity is fairly low, and it would be possible to pair a precision silicon position detector with a timing detector with larger pads.   As long as the two are close together and the occupancy in the timing detector is fairly low, this is as effective as an all-silicon system.   

\section{Conclusions}

Timing can be used to improve the momentum resolution of silicon detectors, and thereby partially compensate for the loss of resolution due to multiple scattering by low-momentum detectors.  The resolution depends on the size of the silicon detector and on the magnetic field. It improves linearly with increasing magnetic field and faster than linearly with increasing detector radius  With a moderate sized tracking detector ($>1$ m radius) and a multi-Tesla magnetic field, useful resolution can be obtained up to $p \approx 1$ GeV/c, bridging the gap to the region where multiple scattering is less important for conventional tracking.  With solenoidal magnetic fields, this approach is most effective for tracks at mid-rapidity; as $|\eta|$ rises, the resolution worsens.  

This approach could be of use at an electron-ion collider, where it is important to accurately reconstruct the entire event.

The technique might also be of use at the Large Hadron Collider, where the ATLAS, CMS and LHCb collaborations are all planning on instrumenting their detectors with timing, for pileup rejection  \cite{Cartiglia:2019fkq,Staszewski:2019yek};  These timing detectors might also be employed to improve the momentum resolution for lower momentum tracks.  field.    Timing for tracking was considered in Ref. \cite{Garzon:2011zza}.  However, that paper only considered straight-line tracks, where velocity was used to reduce tracking confusion. 

\acknowledgments

The author thanks Michael Lomnitz for preparing Figure 1 and many discussions, Yuan Mei for useful conversations, and Peter Jacobs for editing suggestions.  This work is supported by U.S. Department of Energy, Office of Science, Office of Nuclear Physics, under contract number DE-AC02-05CH11231.

\end{document}